\documentclass[11pt,a4paper]{article}
\usepackage[in,plain]{fullpage}
\usepackage[centertags]{amsmath}
\usepackage{amssymb}
\usepackage{epsfig}
\renewcommand{\today}{May 1998}
\begin{document}
\begin{titlepage}
\null
\vspace{5mm}
\begin{flushright}
\begin{tabular}{l}
UWThPh-1998-21\\
DFTT 19/98\\
\today
\end{tabular}
\end{flushright}
\vspace{3mm}
\begin{center}
\Large
\textbf{Neutrino Masses and Mixing\\
in the Light of Experimental Data\footnote{
Lecture given by W.G. at the 5th Workshop on High Energy Physics
Phenomenology, January, 12-26, 1998, IUCAA, Pune, India}}\\[5mm]
\normalsize
S.M. Bilenky\\
Joint Institute for Nuclear Research, Dubna, Russia, and\\
INFN, Sezione di Torino,
Via P. Giuria 1, I--10125 Torino, Italy\\[3mm]
C. Giunti\\
INFN, Sezione di Torino, and Dipartimento di Fisica Teorica,
Universit\`a di Torino,\\
Via P. Giuria 1, I--10125 Torino, Italy\\[3mm]
and\\[3mm]
W. Grimus\\
Institute for Theoretical Physics, University of Vienna,\\
Boltzmanngasse 5, A--1090 Vienna, Austria\\
\vspace{10mm}
\textbf{Abstract}\\[3mm]
\begin{minipage}{0.8\textwidth}
All the possible schemes of neutrino mixing
with four massive neutrinos
inspired by the existing experimental indications
in favour of neutrino mixing
are considered.
It is shown that the scheme with a neutrino
mass hierarchy 
is not compatible with
the experimental results, likewise
all other schemes with the masses of three
neutrinos close together and the fourth mass separated by a gap needed
to incorporate the LSND neutrino oscillations.
Only two schemes
with two pairs of neutrinos with close masses
separated by this gap
of the order of 1 eV
are in agreement with the results
of all experiments.
We carefully examine the arguments leading to this conclusion and also
discuss experimental consequences of the two favoured neutrino schemes.
\end{minipage}
\end{center}
\end{titlepage}

\section{Indications in favour of neutrino oscillations}

\subsection{Notation}

Neutrino masses and neutrino mixing are natural phenomena in
gauge theories extending the Standard Model 
(see, for example, Ref.\cite{gauge}). However, for the time being
masses and mixing angles cannot be predicted on theoretical grounds
and they are the central subject of the experimental activity in the field
of neutrino physics.

In the general discussion, we assume that there are $n$
neutrino fields with definite flavours and that 
neutrino mixing is described by a $n \times n$ unitary mixing matrix
$U$ such that
\begin{equation}
\nu_{{\alpha}L}
=
\sum_{j=1}^{n}
U_{{\alpha}j}
\,
\nu_{jL}
\qquad
(\alpha=e,\mu,\tau,s_1,\dots,s_{n-3}) \,.
\label{mixing}
\end{equation}
Note that the neutrino fields $\nu_{\alpha L}$
other than the three active neutrino flavour fields 
$\nu_{e L}$, $\nu_{\mu L}$, $\nu_{\tau L}$
must be sterile to comply with the result of the LEP measurement of
the number of neutrino flavours.
The fields $\nu_{jL}$ ($j=1,\dots,n$) are the left-handed components
of neutrino fields with definite mass $m_j$. We assume the ordering
$m_1 \leq m_2 \leq \ldots \leq m_n$ for the neutrino masses. In
Eq.(\ref{mixing}) and in the following discussion of neutrino
oscillations it does not matter if the neutrinos are of Dirac or Majorana
type. One should only keep in mind that different types cannot mix.

The most striking feature of neutrino masses and mixing is the
quantum-mechanical effect of neutrino oscillations 
\cite{BP78-87}. The probability
of the transition $ \nu_{\alpha} \rightarrow \nu_{\beta} $
is given by
\begin{equation} \label{prob}
P_{\nu_{\alpha}\rightarrow\nu_\beta} =
\left|
\sum_{j=1}^n
U_{\beta j}
U_{\alpha j}^*
\exp\!\left(-i\frac{\Delta{m}^{2}_{j1} L}{2p}\right) 
\right|^2
\end{equation}
where $\Delta{m}^{2}_{j1} \equiv m^2_j - m^2_1$, $L$ is the distance
between source and detector and $p$ is the neutrino
momentum. Eq.(\ref{prob}) is valid for $p^2 \gg m^2_j$
($j=1,\ldots,n$).\footnote{There are additional conditions depending
on the neutrino production and detection processes which must hold for
the validity of Eq.(\ref{prob}). See, e.g., Ref.\cite{osc} and
references therein.} Evidently,
from neutrino oscillation experiments only differences of squares of
neutrino masses can be determined. The probability for 
$\bar\nu_\alpha\to\bar\nu_\beta$ transitions is obtaind from
Eq.(\ref{prob}) by the substitution $U \to U^*$.

\subsection{Indications in favour of neutrino masses and mixing}

At present, indications that neutrinos are
massive and mixed
have been found in
solar neutrino experiments
(Homestake \cite{Homestake},
Kamiokande \cite{Kam-sun},
GALLEX \cite{GALLEX},
SAGE \cite{SAGE}
and
Super-Kamiokande \cite{SK-sun,SK}),
in atmospheric neutrino experiments
(Kamiokande \cite{Kam-atm},
IMB \cite{IMB},
Soudan \cite{Soudan}
and
Super-Kamiokande \cite{SK-atm,SK})
and
in the LSND experiment \cite{LSND}.
From the analyses of the data of these experiments
in terms of neutrino oscillations
it follows that there are three different scales
of neutrino mass-squared differences:
\begin{itemize}
\item \textbf{Solar neutrino deficit:} Interpreted as effect of
neutrino oscillations the relevant value of the mass-squared difference is
determined as
\begin{equation}
\Delta{m}^2_{\mathrm{sun}}
\sim
10^{-5} \, \mathrm{eV}^2
\, (\mbox{MSW})
\quad \mbox{or} \quad
\Delta{m}^2_{\mathrm{sun}}
\sim
10^{-10} \, \mathrm{eV}^2
\, (\mbox{vac. osc.})
\quad
\mbox{\cite{HL97,FLM97}} \,.
\label{DMsun}
\end{equation}
The two possibilities for
$\Delta{m}^2_{\mathrm{sun}}$
correspond,
respectively,
to the
MSW \cite{MSW}
and
to the
vacuum oscillation
solutions of the solar neutrino problem.
\item \textbf{Atmospheric neutrino anomaly:} Interpreted as effect of
neutrino oscillations, the zenith angle dependence of the atmospheric
neutrino anomaly \cite{Kam-atm,SK-atm,SK} gives
\begin{equation}
\Delta{m}^2_{\mathrm{atm}}
\sim
5 \times 10^{-3} \, \mathrm{eV}^2
\quad
\mbox{\cite{Valencia}}
\,.
\label{DMatm}
\end{equation}
\item \textbf{LSND experiment:} The evidence for $\bar\nu_\mu \to \bar\nu_e$
oscillations in this experiment leads to 
\begin{equation}
\Delta{m}^2_{\mathrm{SBL}} \sim 1 \, \mathrm{eV}^2
\quad
\mbox{\cite{LSND}}
\label{DMlsnd}
\end{equation}
where
$\Delta{m}^2_{\mathrm{SBL}}$
is the neutrino mass-squared difference
relevant for short-baseline (SBL) experiments.
\end{itemize}

Thus,
at least four light neutrinos with definite masses
must exist in nature
in order to accommodate the results of all neutrino oscillation
experiments. Denoting by $\delta m^2$ a generic neutrino mass-squared
difference we can summarize the discussion in the following way:
\begin{quote}
$\diamondsuit$ 3 different scales of $\delta m^2
\quad \Rightarrow \quad$4 neutrinos (or more).
\end{quote}
Therefore there exists at least one non-interacting sterile 
neutrino
\cite{four,BGKP,BGG96,OY96,BGG97a,BGG97b,Barger-Gibbons}.

However,
we must also take into account
the fact that
in several short-baseline experiments
neutrino oscillations were not observed.
The results of these experiments
allow to exclude
large regions in the space of the neutrino oscillation
parameters. This will be done in the next section.

The plan of this report is as follows. In section II we extensively
discuss SBL neutrino oscillations for an arbitrary number of
neutrinos. In section III we argue that a 4-neutrino mass hierarchy is
disfavoured by the experimental data. Thereby, solar and atmospheric
neutrino flux data play a crucial role. In section IV we introduce the
two 4-neutrino mass and mixing schemes favoured by all 
neutrino oscillation experiments. We discuss possibilities to check these
schemes in long-baseline (LBL) neutrino oscillation experiments in
section V. Our conclusions are presented in section VI.

\section{SBL experiments}

\subsection{The oscillation phase}

As a guideline, SBL neutrino oscillation experiments are 
sensitive to mass-squared differences $\delta m^2 > 0.1$ eV$^2$. 
A generic oscillation phase is given by
\begin{equation}\label{phase}
\frac{\delta m^2 L}{2p} \simeq 2.53 \times 
\left( \frac{\delta m^2}{1\, \mbox{eV}^2} \right)
\left( \frac{p}{1\, \mbox{MeV}} \right)^{-1}
\left( \frac{L}{1\, \mbox{m}} \right) \,.
\end{equation}
Distinguishing reactor and accelerator experiments and assuming that
experiments are roughly sensitive to phases (\ref{phase}) around 0.1
or larger we get the following conditions from  $\delta m^2 > 0.1$ eV$^2$:
\begin{itemize}
\item Reactors: $p \sim 1$ MeV and therefore $L \gtrsim 10$ m.
\item Accelerators: $L \gtrsim 10^3 \,\mbox{m} \, \times
(p/1\,\mbox{GeV})$.
\end{itemize}

\subsection{Basic assumption and formalism}

We will make the following basic assumption in the further discussion
in this report: 
\begin{quote}
$\diamondsuit$ A 
single $\delta m^2$ is relevant in SBL neutrino experiments.
\end{quote}
In accordance with Eq.(\ref{DMlsnd}) we denote this $\delta m^2$ by
$\Delta m^2_{\mbox{\scriptsize SBL}}$.

As a consequence of this assumption
the neutrino mass
spectrum consists of two groups of
close masses,
separated by a mass difference in the eV range.
Denoting the neutrinos of the two groups by
$ \nu_1, \ldots , \nu_r $
and
$ \nu_{r+1}, \ldots , \nu_n $,
respectively, the mass spectrum looks like
\begin{equation}
m^2_1 \leq \ldots  \leq m^2_r \ll
m^2_{r+1} \leq \ldots \leq m^2_n
\end{equation}
such that
\begin{equation}\label{sbl}
\begin{array}{lcccc}
\Delta m^2_{kj} \ll \Delta m^2_{\mbox{\scriptsize SBL}} 
& \mbox{ for } & 1 \leq j < k \leq r & 
\mbox{ and } & r+1 \leq j < k \leq n, \\[2mm]
\Delta m^2_{kj} \simeq \Delta m^2_{\mbox{\scriptsize SBL}} & 
\mbox{ for } & 1 \leq j \leq r & 
\mbox{ and } & r+1 \leq k \leq n
\end{array}
\end{equation}
for the purpose of the SBL formalism. In Eq.(\ref{sbl}) we have used
the notation $\Delta m^2_{kj} \equiv m^2_k-m^2_j$.
Eq.(\ref{prob}) together with Eq.(\ref{sbl}) gives the SBL transition
probability 
\begin{equation} \label{sblprob}
P^{(\mbox{\scriptsize SBL})}_{\nu_{\alpha}\rightarrow\nu_\beta} =
\left|
\sum_{j=1}^r
U_{\beta j}
U_{\alpha j}^* +
\exp\!\left(-i\frac{\Delta m^2_{\mbox{\scriptsize SBL}} L}{2p}\right) 
\sum_{j=r+1}^n
U_{\beta j}
U_{\alpha j}^* \right|^2 \,.
\end{equation}

For the probability of the transition
$\nu_{\alpha}\rightarrow\nu_{\beta}$
($\alpha\neq\beta$) we obtain from Eq.(\ref{sblprob}) 
\begin{equation} \label{Pa b}
P^{(\mbox{\scriptsize SBL})}_{\nu_{\alpha}\rightarrow\nu_{\beta}}
=
\frac{1}{2}
A_{\alpha;\beta}
\left( 1 - \cos \frac{\Delta m^2_{\mbox{\scriptsize SBL}} L}{2p} \right)
\end{equation}
where the oscillation amplitude
$A_{\alpha;\beta}$
is given by
\begin{equation} \label{Aab}
A_{\alpha;\beta}  =
4 \left| \sum_{j \geq r+1} U_{\beta j} U_{\alpha j}^* \right|^2
=
4 \left| \sum_{j \leq r}
U_{\beta j} U_{\alpha j}^* \right|^2
\,.
\end{equation}
The second equality sign in this equation follows from the unitarity of $U$.
Furthermore, the oscillation amplitude $A_{\alpha;\beta}$ fulfills
the condition $A_{\alpha;\beta} = A_{\beta;\alpha} \leq 1$.
The second part of this equation is a consequence of 
the Cauchy--Schwarz inequality and the unitarity of the mixing matrix.
The survival probability of $\nu_{\alpha}$ is
calculated as
\begin{equation} \label{Pa}
P^{(\mbox{\scriptsize SBL})}_{\nu_{\alpha}\rightarrow\nu_{\alpha}}
=
1 - \sum_{\beta\neq\alpha}
P_{\nu_{\alpha}\rightarrow\nu_\beta}
=
1 - \frac{1}{2}
B_{\alpha;\alpha}
\left(1 - \cos \frac{\Delta{m}^2_{\mbox{\scriptsize SBL}} L}{2p} \right)
\end{equation}
with the survivial amplitude
\begin{equation} \label{Ba}
B_{\alpha;\alpha} =
4 \left( \sum_{j \geq r+1} |U_{\alpha j}|^2 \right)
\left( 1 - \sum_{j \geq r+1} |U_{\alpha j}|^2 \right)
= 4 \left( \sum_{j \leq r} |U_{\alpha j}|^2 \right)
\left( 1 - \sum_{j \leq r} |U_{\alpha j}|^2 \right) \,.
\end{equation}
Conservation of probability gives the important relation
\begin{equation}\label{BSA}
B_{\alpha;\alpha}
=
\sum_{\beta \neq \alpha}
A_{\alpha;\beta} \leq 1 \,.
\end{equation}

The expressions (\ref{Pa b}) and (\ref{Pa})
describe the transitions between all possible
neutrino states,
whether active or sterile.
Let us stress that with the basic assumption in the beginning of this
subsection the oscillations in
all channels are characterized
by the same oscillation length
$l_{\mbox{\scriptsize osc}} = 4 \pi p / \Delta
m^2_{\mbox{\scriptsize SBL}}$. Furthermore, the
substitution $U \to U^*$ in the amplitudes (\ref{Aab}) and (\ref{Ba})
does not change them and therefore it ensues from the basic SBL assumption
that the probabilities (\ref{Pa b}) and (\ref{Pa}) hold for
antineutrinos as well and hence there is no CP violation in SBL
neutrino oscillations.

The oscillation probabilities (\ref{Pa b}) and (\ref{Pa}) look like
2-flavour probabilities. Defining  
$\sin^2 2\theta_{\alpha\beta} \equiv A_{\alpha;\beta}$,
$\sin^2 2\theta_{\alpha} \equiv B_{\alpha;\alpha}$ and
$\sin^2 2\theta_{\beta} \equiv B_{\beta;\beta}$ for $\alpha \neq \beta$,
the resemblance is even more striking. It means that the basic SBL
assumption allows to use the 2-flavour oscillation formulas in SBL
experiments.
However, genuine 2-flavour 
$\nu_\alpha \leftrightarrow \nu_\beta$ neutrino oscillations
are characterized by a single mixing angle given by 
$\theta_{\alpha\beta} = \theta_\alpha = \theta_\beta$.

\subsection{Disappearance experiments}

For the two flavours $\alpha = e$ and $\mu$ results of disappearence
experiments are available. We will use the 90\% exclusion plots of the
Bugey reactor experiment \cite{Bugey95} for 
$\bar\nu_e\to\bar\nu_e$ disappearance
and the 90\% exclusion plots of the CDHS \cite{CDHS84} and CCFR
\cite{CCFR84} accelerator experiments for $\nu_\mu\to\nu_\mu$
disappearance. Since no neutrino disappearance has been seen there are
upper bounds $B_{\alpha;\alpha}^0$ on the disappearance amplitudes for
$\alpha = e, \mu$.
These experimental bounds are functions of
$\Delta m^2_{\mbox{\scriptsize SBL}}$. It follows that
\begin{equation}\label{ca}
B_{\alpha,\alpha} = 4\, c_\alpha (1-c_\alpha) \leq B_{\alpha;\alpha}^0
\quad \mbox{with} \quad
c_\alpha \equiv \sum_{j=1}^r |U_{\alpha j}|^2
\end{equation}
and therefore \cite{BBGK}
\begin{equation}
c_\alpha \leq a^0_\alpha \quad \mbox{or} \quad c_\alpha \geq
1-a^0_\alpha \quad \mbox{with} \quad \label{a0}
a^{0}_{\alpha} \equiv \frac{1}{2}
\left(1-\sqrt{1-B_{\alpha;\alpha}^{0}}\,\right)
\,.
\end{equation}
Eq.(\ref{a0}) shows that $a^0_\alpha \leq 1/2$. In Fig.\ref{a0plot} the bounds
$a^0_e$ and $a^0_\mu$ are plotted as functions of
$\Delta m^2_{\mbox{\scriptsize SBL}}$ in the wide range
\begin{equation}\label{wr}
10^{-1} \leq \Delta m^2_{\mbox{\scriptsize SBL}} \leq 
10^3 \mbox{ eV}^2 \,.
\end{equation}
In this range $a^0_e$ is small ($a^0_e \lesssim 4 \times 10^{-2}$) and
$a^0_\mu \lesssim 10^{-1}$ for
$\Delta m^2_{\mbox{\scriptsize SBL}} \gtrsim 0.5$ eV$^2$. This means
that in the $c_e$--$c_\mu$ unit square for every 
$\Delta m^2_{\mbox{\scriptsize SBL}}$ we can distinguish four allowed 
regions according to $c_\alpha \leq a^0_\alpha$ or 
$c_\alpha \geq a^0_\alpha$ (see Fig.\ref{square}).
\renewcommand{\arraystretch}{1.5}
\begin{table}[t!]
\begin{tabular*}{\textwidth}{@{\extracolsep{\fill}}cc}
\begin{minipage}{0.49\linewidth}
\begin{center}
\mbox{\epsfig{file=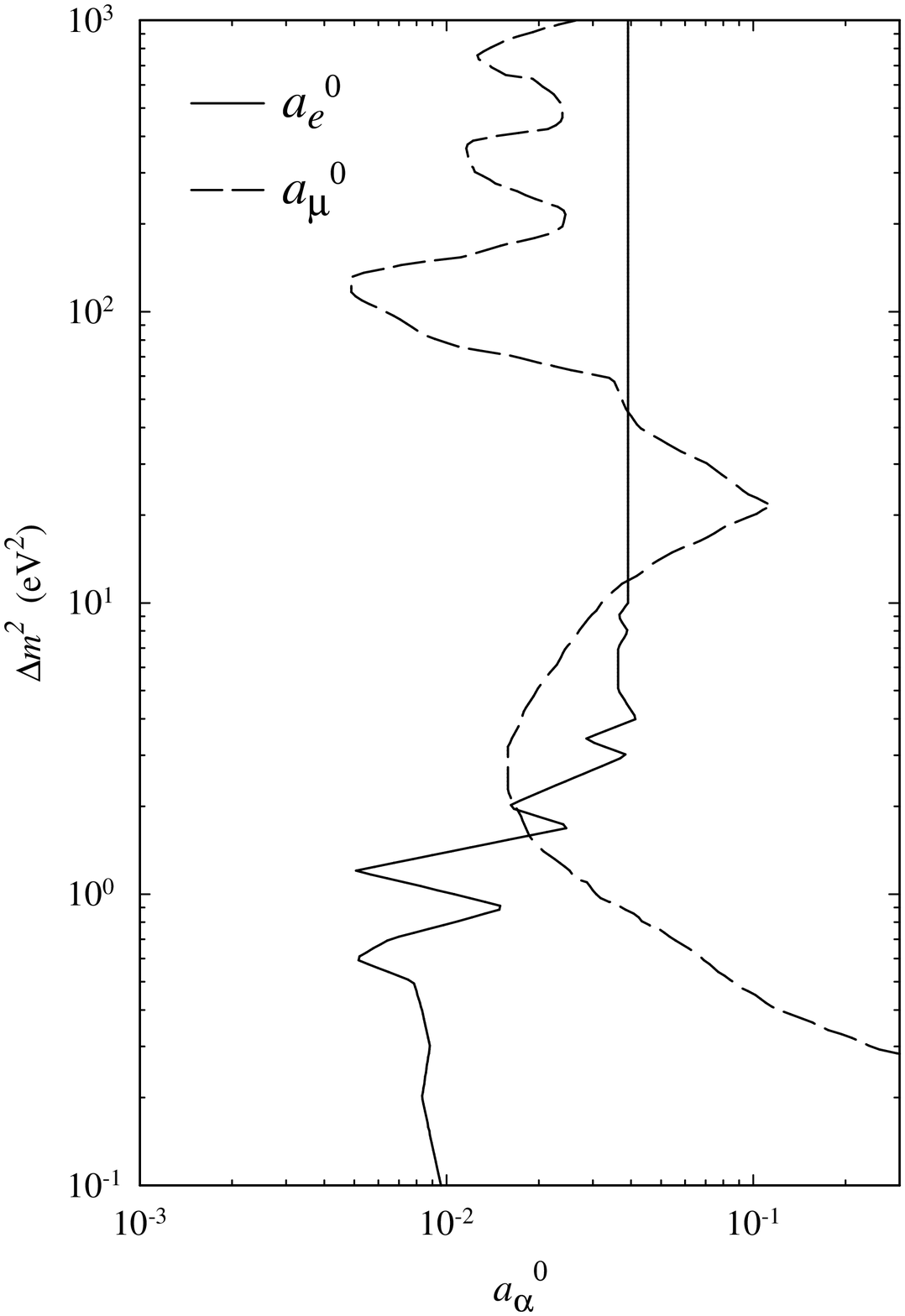,width=0.95\linewidth}}
\end{center}
\end{minipage}
&
\begin{minipage}{0.49\linewidth}
\begin{center}
\setlength{\unitlength}{1.4cm}
\begin{picture}(6,6)(-0.8,0.05)
\put(0,0){\vector(1,0){4.3}}
\put(0,0){\vector(0,1){4.3}}
\put(0,4){\line(1,0){4}}
\put(4,0){\line(0,1){4}}
\put(4.5,0){\makebox(0,0)[l]{$c_e$}}
\put(0,4.5){\makebox(0,0)[b]{$c_\mu$}}
\put(0,1){\line(1,0){0.6}}
\put(0.6,0){\line(0,1){1}}
\put(0,3){\line(1,0){0.6}}
\put(0.6,4){\line(0,-1){1}}
\put(4,1){\line(-1,0){0.6}}
\put(3.4,0){\line(0,1){1}}
\put(4,3){\line(-1,0){0.6}}
\put(3.4,4){\line(0,-1){1}}
\put(0.3,0.5){\makebox(0,0){I}}
\put(3.7,0.5){\makebox(0,0){II}}
\put(3.7,3.5){\makebox(0,0){III}}
\put(0.3,3.5){\makebox(0,0){IV}}
\put(0.6,-0.1){\makebox(0,0)[t]{$a^0_e$}}
\put(3.4,-0.1){\makebox(0,0)[t]{$1-a^0_e$}}
\put(-0.1,1){\makebox(0,0)[r]{$a^0_\mu$}}
\put(-0.1,2.9){\makebox(0,0)[r]{$1-a^0_\mu$}}
\put(-0.1,-0.15){\makebox(0,0)[tr]{0}}
\put(4,-0.15){\makebox(0,0)[t]{1}}
\put(-0.1,4){\makebox(0,0)[r]{1}}
\end{picture}
\end{center}
\end{minipage}
\\
\refstepcounter{figure}
\label{a0plot}                 
Figure \ref{a0plot}
&
\refstepcounter{figure}
\label{square}                 
Figure \ref{square}
\end{tabular*}
\null \vspace{-0.5cm} \null
\end{table}

\subsection{The
$\stackrel{\scriptscriptstyle (-)}{\nu}_{\hskip-3pt \mu} 
\to
\stackrel{\scriptscriptstyle (-)}{\nu}_{\hskip-3pt e}$ 
transition in SBL experiments}

Considering the amplitude $A_{\mu;e}$,
with the help of the Cauchy--Schwarz inequality
we obtain from Eq.(\ref{Aab})
\begin{equation}\label{Amin}
A_{\mu;e} \leq 4\, \min \left[ c_e c_\mu, (1-c_e)(1-c_\mu)
\right] \,.
\end{equation}
Therefore, 
we immediately see that
\begin{equation}\label{mebound}
A_{\mu;e} \leq 4\, a^0_e a^0_\mu \quad \mbox{in regions I and III}.
\end{equation}

In Fig.\ref{emu} 
the result of the LSND experiment \cite{LSND} for the amplitude
$A_{\mu;e}$ is shown with 90\% CL boundaries (shaded areas). 
All other experiments measuring this amplitude have obtained 
upper bounds \cite{BNLE734,BNLE776,KARMEN,CCFR97}.
In addition, the upper bound $B^0_{e;e}$ on the 
$\bar\nu_e\to\bar\nu_e$ survivial amplitude of Bugey \cite{Bugey95} is
indicated by the solid line in Fig.\ref{emu}
since the unitarity relation (\ref{BSA}) gives
$A_{\mu;e} \leq B^0_{e;e}$.
Finally, the curve passing through the circles represents the bound
(\ref{mebound}). Inspecting Fig.\ref{emu} we come to the following conclusion:
\begin{quote}
$\diamondsuit$ Regions I and III are not compatible
with the positive result of LSND indicating $\bar\nu_\mu\to\bar\nu_e$
oscillations and the negative results of all other SBL experiments.
\end{quote}
Furthermore, it can be read off from Fig.3 that
\begin{equation}
0.27 \, \mathrm{eV}^2
\lesssim
\Delta{m}^2_{\mathrm{SBL}}
\lesssim
2.2 \, \mathrm{eV}^2
\label{LSNDrange}
\end{equation}
is the favoured range for the SBL mass-squared difference.
In this range $a^0_\mu \lesssim 0.3$ holds.
Let us further mention that for $r=1$ region III is already ruled out
by the unitarity of the mixing matrix. The same is valid for $r=n-1$
and region I.

\section{The 4-neutrino mass hierarchy is disfavoured}

In the case of
a neutrino mass hierarchy,
$ m_1 \ll m_2 \ll m_3 \ll m_4 $,
the mass-squared differences $\Delta{m}^{2}_{21}$ and 
$\Delta{m}^{2}_{32}$ are relevant
for the suppression of the flux of solar neutrinos and for the
atmospheric neutrino anomaly, respectively.
This case
corresponds to $n=4$ and $r=3$ (see the formalism in subsection 2.2)
with
$c_\alpha = \sum_{j=1}^3 |U_{\alpha j}|^2$.
We only have to consider regions II and IV.

We will now take into account information from the solar neutrino
anomaly assuming that it is solved by neutrino oscillations. From the
fact that the 4th column vector in $U$ pertaining to $m_4$ is not
affected by solar neutrino oscillations we obtain a lower bound on 
the average survival probability of solar
neutrinos given by
(see Refs.\cite{SS92,BGKP})
\begin{equation} \label{Psol}
P^{\odot}_{\nu_e\rightarrow\nu_e} \geq |U_{e4}|^4 \,.
\end{equation}
In region IV we have $c_e \leq a^0_e$ or $|U_{e4}|^2 \geq 1-a^0_e$
and therefore 
$ P^{\odot}_{\nu_e\rightarrow\nu_e} \gtrsim 0.92 $
holds for all solar neutrino energies.
Such a
large lower bound is not compatible with the solar neutrino
data and we conclude:
\begin{quote}
$\diamondsuit$ For
a 4-neutrino mass hierarchy region IV is not compatible with the
solar neutrino data.
\end{quote}
Let us mention that inequality (\ref{Psol}) is not completely
exact. In the solar neutrino problem the matter background is
important and it enters the total Hamiltonian for neutrino
propagation. Nevertheless, to very good accuracy the largest
eigenvalue of the Hamiltonian is given by 
$E_4 \simeq m^2_4/2p$ with eigenvector
$v_4 \simeq (U_{\alpha 4})$ and corrections to this are of order
$a_{CC}/\Delta m^2_{\mbox{\scriptsize SBL}} \sim 10^{-5}$ 
where $a_{CC} = 2\sqrt{2} G_F N_e p$,
$N_e$ denotes the electron number density in the sun and in the solar
core $a_{CC} \sim 10^{-5}$ eV$^2$. Furthermore, the evolution of $v_4$
in solar matter is adiabatic to an even better accuracy. Thus
Eq.(\ref{Psol}) is accurate for our purpose.

\begin{table}[t!]
\begin{tabular*}{\textwidth}{@{\extracolsep{\fill}}cc}
\begin{minipage}{0.49\linewidth}
\begin{center}
\mbox{\epsfig{file=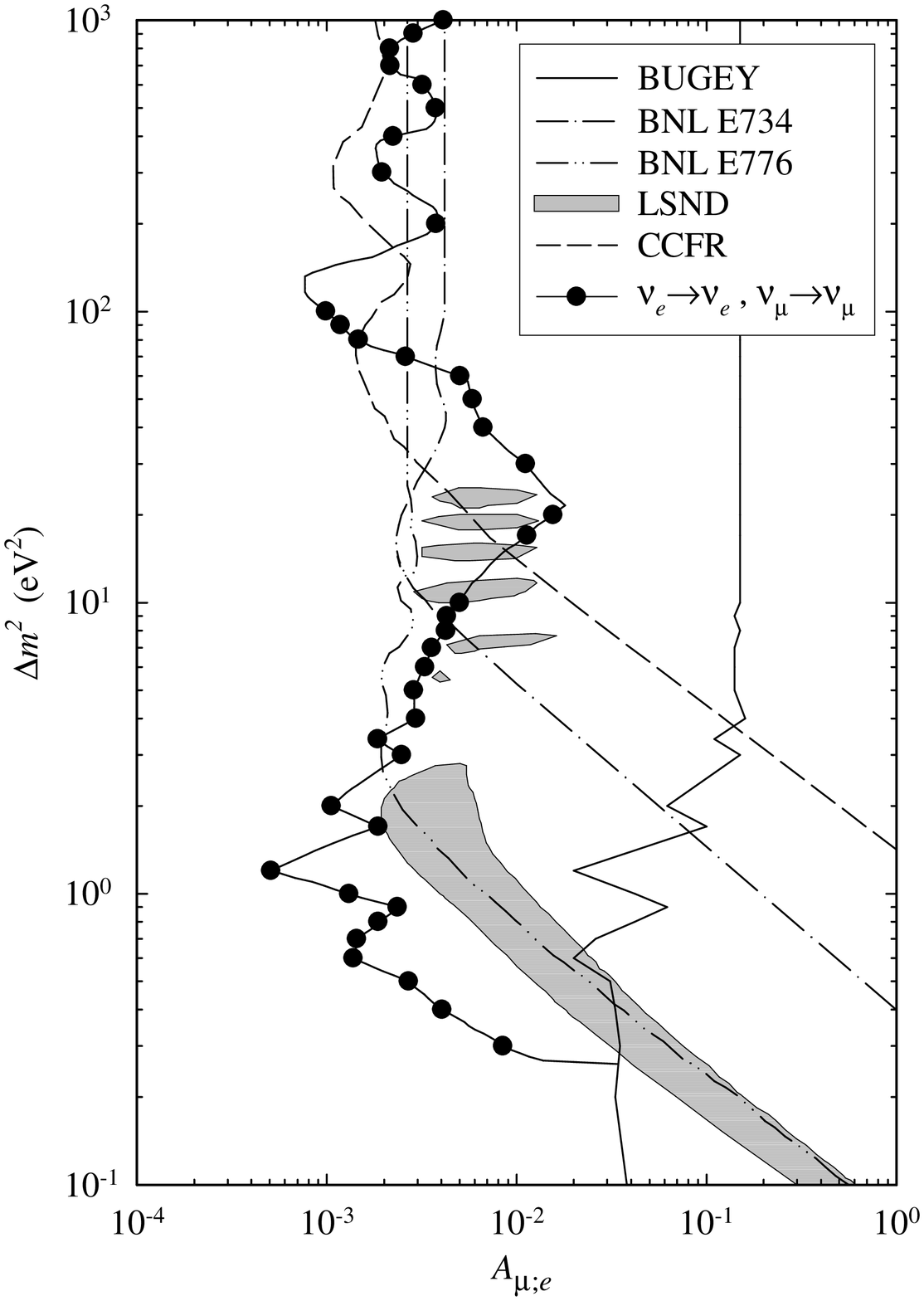,width=0.95\linewidth}}
\end{center}
\end{minipage}
&
\begin{minipage}{0.49\linewidth}
\begin{center}
\mbox{\epsfig{file=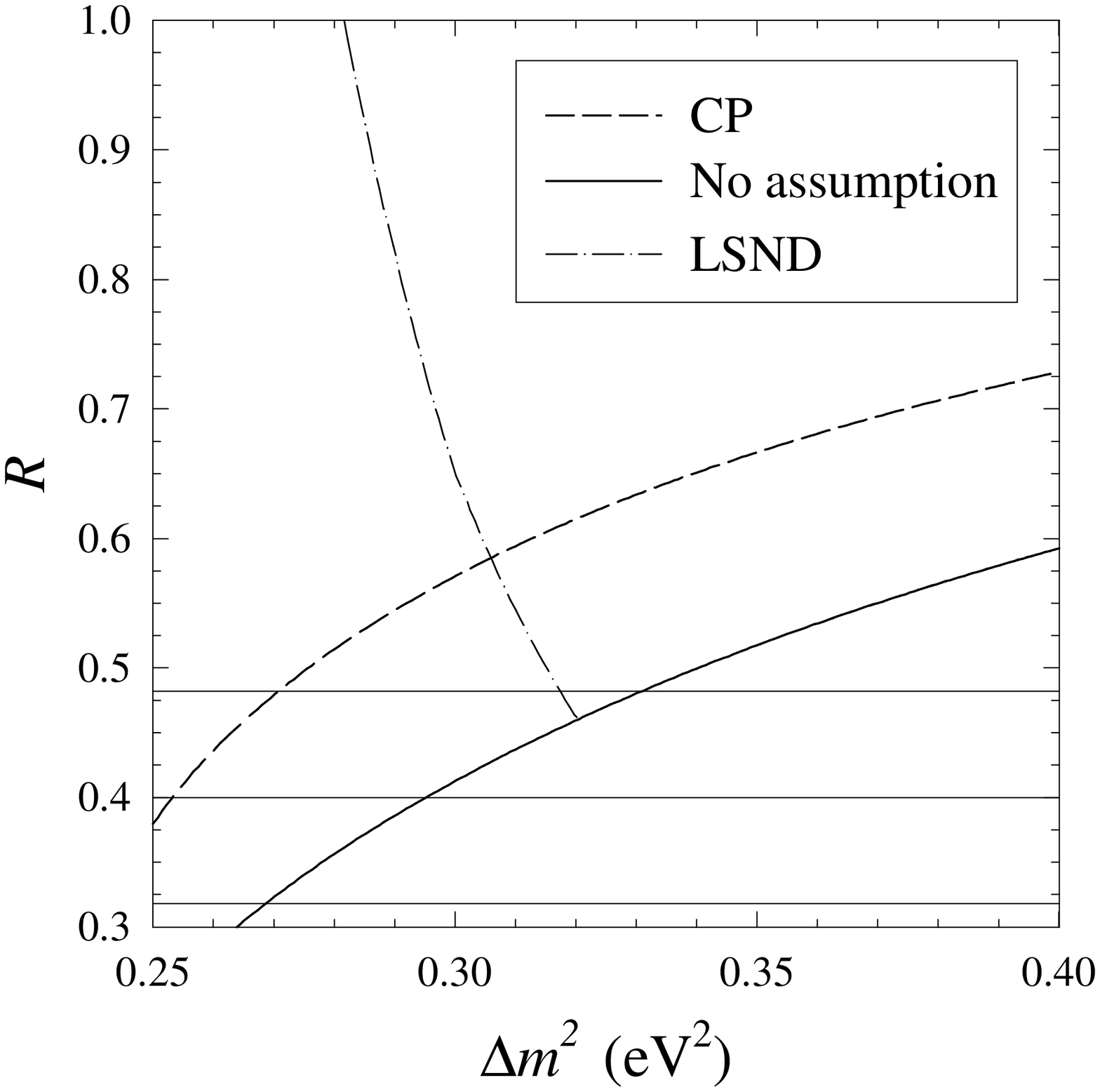,width=0.95\linewidth}}
\end{center}
\end{minipage}
\\
\refstepcounter{figure}
\label{emu}                 
Figure \ref{emu}
&
\refstepcounter{figure}
\label{atmf}                 
Figure \ref{atmf}
\end{tabular*}
\null \vspace{-0.5cm} \null
\end{table}
It remains to discuss region II. To this end we consider the
atmospheric neutrino anomaly which is expressed through the deviation of
the double ratio
\begin{equation}\label{R}
R = 
\frac{(\mu/e)_{\mbox{\scriptsize data}}}{(\mu/e)_{\mbox{\scriptsize MC}}}
=
\frac{P^{\mathrm{atm}}_{\nu_\mu\to\nu_\mu} +
r^{-1} P^{\mathrm{atm}}_{\nu_e\to\nu_\mu}}{
P^{\mathrm{atm}}_{\nu_e\to\nu_e} +
r P^{\mathrm{atm}}_{\nu_\mu\to\nu_e}}
\end{equation}
from 1. In Eq.(\ref{R}) 
$(\mu/e)_{\mbox{\scriptsize MC}} \equiv r$ is the ratio of
muon and electron events without neutrino oscillations. It is
obtained by a Monte Carlo calculation which gives
$r \simeq 1.57$ for sub-GeV events. For atmospheric neutrinos matter effects
are non-negligible.
Analogously to Eq.(\ref{Psol}) we have the lower bound
\begin{equation}\label{Patm}
P^{\mathrm{atm}}_{\nu_\mu\to\nu_\mu} \geq |U_{\mu 4}|^4 \,.
\end{equation}

Let us assume for the moment that
$
P^{\mathrm{atm}}_{\nu_e\to\nu_\mu}=P^{\mathrm{atm}}_{\nu_\mu\to\nu_e} 
$.
This is the case if CP is conserved or if the oscillating parts in the
probabilities occurring in Eq.(\ref{Patm}) drop out because of
averaging processes involving neutrino energy and distance between
source and detector. Then it is easily shown by Eqs.(\ref{R}) and
(\ref{Patm})
that \cite{BGG96}
\begin{equation}\label{P=Pineq}
R \geq P^{\mathrm{atm}}_{\nu_\mu\to\nu_\mu} \geq (1-c_\mu)^2 
\end{equation}
for all energy ranges and zenith angle bins.
In this case in region II we obtain
\begin{equation}\label{P=Pbound}
R \geq (1-a^0_\mu)^2 \,.
\end{equation}

The assumption
$
P^{\mathrm{atm}}_{\nu_e\to\nu_\mu}=P^{\mathrm{atm}}_{\nu_\mu\to\nu_e} 
$  
is not fully satisfactory because it is not
clear if or how well it is fulfilled. Let us therefore dispense with
it now.
The evolution of oscillation probabilities with a matter background 
has the general form \cite{SS92,BGKP} 
\begin{equation}\label{PB}
P_{\nu_\alpha \to \nu_\beta}(x_1,x_0) =
\left| U(x_1)_{\beta k} B_{kj} U(x_0)^*_{\alpha j} \right|^2
\end{equation}
where $B$ is a unitary matrix and $U(x)$ diagonalizes the
Hamiltonian for neutrino propagation in matter at the location $x$.
Note that Eq.(\ref{PB}) is the generalization of Eq.(\ref{prob})
referring to vacuum oscillations where
$B$ has only diagonal elements given by 
$\exp (-i \Delta m^2_{j1}(x_1-x_0)/2p)$ and $U(x_0)=U(x_1)=U$.
Because $\Delta m^2_{\mbox{\scriptsize SBL}} \gg a_{CC},
\: \Delta m^2_{\mathrm{atm}}, \: \Delta m^2_{\mathrm{sun}}$, the
matrix $B$ decomposes approximately into a 3 $\times$ 3 and a 1
$\times$ 1 block and therefore (see the discussion after Eq.(\ref{Psol}))
\begin{equation}\label{p10}
P_{\nu_\alpha \to \nu_\beta}(x_1,x_0) \simeq
\sum_{j,k=1}^3 \left| U(x_1)_{\beta k} B_{kj} U(x_0)^*_{\alpha j} \right|^2
+ |U_{\beta 4}|^2 |U_{\alpha 4}|^2 \,.
\end{equation}
This consideration leads to
\begin{equation}\label{general}
R \geq \frac{(1-c_\mu)^2 + r^{-1} (1-c_e)(1-c_\mu)}{
             c^2_e + (1-c_e)^2 + 
             r [c_e c_\mu + (1-c_e)(1-c_\mu) ]} 
\geq \frac{(1-c_\mu)^2}{1+rc_\mu} \,.
\end{equation}
For $c_e \geq 1-a^0_e$, the central expression of Eq.(\ref{general}) has 
the minimum with respect to $c_e$ at $c_e=1$.
This explains the second part of the inequality.
Eq.(\ref{general}) represents a general bound valid for all
energy ranges and zenith angles, whether assumption 
$
P^{\mathrm{atm}}_{\nu_e\to\nu_\mu}=P^{\mathrm{atm}}_{\nu_\mu\to\nu_e} 
$
is fulfilled or not. Its right-hand side is a
decreasing function in $c_\mu$ and therefore in region II we arrive at
\begin{equation}\label{generala0}
R \geq \frac{(1-a^0_\mu)^2}{1+ra^0_\mu} \,.
\end{equation}
Let us take advantage of the $\cos \zeta = -0.8$ bin
($\zeta$ is the zenith angle) of the sub-GeV Superkamiokande
events where $R \lesssim 0.48$ (90\% CL) \cite{kearns}. Here 
$R$ is particularly small. In
Fig.\ref{atmf} the horizontal lines indicate $R$ 
with its 90\% CL interval taken from
Ref.\cite{kearns}, the dashed line represents the bound
(\ref{P=Pbound}) and the solid line the general bound
(\ref{generala0}). Taking into account that the SBL experiments and, in
particular, LSND restrict $\Delta m^2_{\mbox{\scriptsize SBL}}$ to the
range (\ref{LSNDrange}) ($\Delta m^2_{\mbox{\scriptsize SBL}} \gtrsim
0.27$ eV$^2$) we see that the bound (\ref{P=Pbound}) rules out region II. 
However, the general bound (\ref{generala0}) 
it is not tight enough around 
$\Delta m^2_{\mbox{\scriptsize SBL}} \sim 0.3$ eV$^2$ 
to fully exclude region II with a neutrino mass hierarchy because
$a^0_\mu$ gets too large there.

There is a possiblity to improve the bound around 0.3 eV$^2$ in
the following way. For a mass hierarchy we have
$A_{\mu;e} = 4 (1-c_e)(1-c_\mu)$
or 
$c_\mu = 1 - A_{\mu;e}/4(1-c_e) \leq
        1 - A^{\mathrm{min}}_{\mu;e}/4 a^0_e$
where $A^{\mathrm{min}}_{e;\mu}$ is the minimum measured by
LSND. Thus we get
\begin{equation}\label{abarbound}
R \geq \frac{(1-\bar a^0_\mu)^2}{1+r\bar a^0_\mu} \quad \mbox{with}
\quad \bar a^0_\mu 
\equiv \min (a^0_\mu, \frac{A^{\mathrm{min}}_{\mu;e}}{4 a^0_e}) \,.
\end{equation}
The dash-dotted curve in Fig.\ref{atmf} which branches off from the
solid curve corresponds to the part of the lower bound
(\ref{abarbound}) originating from
$A^{\mathrm{min}}_{\mu;e}$. Therefore, comparing 
the lower bounds on $R$ obtained by using 90\% CL data, 
namely the solid and the dash-dotted lines,
with the uppermost
horizontal line which corresponds to the 90\% CL
experimental upper bound on $R$
we see that only a tiny allowed triangle is left in Fig.\ref{atmf}.
Thus we arrive at the conclusion:
\begin{quote}
$\diamondsuit$ With a 4-neutrino mass hierarchy 
region II is strongly disfavoured by
the atmospheric neutrino data and the results of all SBL neutrino
oscillation experiments.
\end{quote}

Let us summarize our findings for a 4-neutrino mass hierarchy:
\begin{itemize}
\item Region I: Excluded by the unitarity of $U$.
\item Region II: Strongly disfavoured by atmospheric neutrino data.
\item Region III: Ruled out by LSND.
\item Region IV: Ruled out by solar neutrino data.
\end{itemize}
It is easy to show that with the arguments presented here all neutrino
mass schemes where three masses are clustered and the fourth one is
separated by the ``LSND gap'' are disfavoured by the present data
\cite{BGG96,OY96}. 

\section{The favoured non-hierarchial 4-neutrino mass spectra}

Now we are left with only 
two possible neutrino mass spectra
in which the four neutrino
masses appear in two pairs
separated by
$ \sim 1 \, \mathrm{eV} $:
\begin{equation} \label{AB}
\mbox{(A)}
\qquad
\underbrace{
\overbrace{m_1 < m_2}^{\mbox{atm}}
\ll
\overbrace{m_3 < m_4}^{\mbox{solar}}
}_{\mbox{LSND}}
\qquad \mbox{and} \qquad
\mbox{(B)}
\qquad
\underbrace{
\overbrace{m_1 < m_2}^{\mbox{solar}}
\ll
\overbrace{m_3 < m_4}^{\mbox{atm}}
}_{\mbox{LSND}}
\;.
\end{equation}
We have to check that these mass spectra are compatible with the
results of all neutrino oscillation experiments.

In schemes A and B the quantities $c_\alpha$ (\ref{ca}) are 
defined with $r=2$.
Clearly, regions I and III (see Fig.\ref{square}) are ruled out by
LSND (see subsection 2.4).
Let us first consider scheme A.
For the survival probability of solar
$\nu_e$'s have \cite{SS92,BGKP}
\begin{equation} \label{sun}
P^{\odot}_{\nu_e\rightarrow\nu_e} = \sum_{i=1,2} |U_{ei}|^4 +
(1-c_e)^2 P^{(3;4)}_{\nu_e\rightarrow\nu_e} 
\end{equation}
where $P^{(3;4)}_{\nu_e\rightarrow\nu_e}$ is the $\nu_e$
survival probability involving $\nu_3$, $\nu_4$ only.
If $ c_e \geq 1-a^{0}_e $,
it follows from Eq.(\ref{sun}) that
the survival
probability $P^{\odot}_{\nu_e\rightarrow\nu_e}$ of solar $\nu_e$'s
practically does not depend
on the neutrino energy
and
$P^{\odot}_{\nu_e\rightarrow\nu_e} \gtrsim 0.5$.
This is disfavoured
by the solar neutrino data
\cite{KP96}. Consequently, regions II and III are ruled out by the solar
neutrino data. This argument does not apply to region IV and one can
easily convince oneself that also the atmospheric neutrino anomaly is
compatible with this region. Furthermore, looking at 
Eq.(\ref{Amin}) we see that this upper bound on $A_{\mu;e}$
is linear in the small quantity
$a^{0}_{e}$ in region IV.
Since
$ a^{0}_{e} \gtrsim 5 \times 10^{-3} $
for all values of
$\Delta m^2_{\mbox{\scriptsize SBL}}$,
in the case of scheme A
the bound (\ref{Amin})
is compatible with the result of the LSND experiment. For scheme B the
analogous arguments lead to region II. Therefore we come to the
conclusion that \cite{BGG96,OY96}
\begin{equation}
\begin{array}{lclccl}
\mbox{Scheme A:}\quad & c_e \leq & a^0_e & \mbox{ and } & c_\mu \geq
& 1-a^0_\mu \,, \\
\mbox{Scheme B:}\quad & c_e \geq & 1-a^0_e & \mbox{ and } & c_\mu \leq
& a^0_\mu \,. \label{schemes}
\end{array}
\end{equation}
Schemes A and B have different consequences for the 
mesurement of the neutrino mass
through the investigation
of the end-point part of the $^3$H $\beta$-spectrum.
From Eq.(\ref{schemes})
it follows that
in the case of scheme A
the neutrino mass that enters in the
usual expression for the $\beta$ spectrum of
$^3$H decay
is approximately equal to
the ``LSND mass'', i.e., $m_{\nu}(^3\mathrm{H}) \simeq m_4$.
If scheme B
is realized in nature
and $m_1$, $m_2$ are very small,
the mass measured in
$^3$H experiments
is at least two order
of magnitude smaller than $m_4$ \cite{BGG96,OY96}.

\section{Checks of the favoured neutrino schemes in LBL experiments}

LBL neutrino oscillation experiments are sensitive to the so-called
``atmospheric $\delta m^2$ range'' of $10^{-2}$--$10^{-3}$ eV$^2$. For
reactor experiments with $p \sim 1$ MeV this amounts to $L \sim 1$
km \cite{CHOOZ,PV} 
whereas in accelerator experiments with $p \sim 1$--10 GeV the
length of the baseline is of order $L \sim 1000$ km 
\cite{KEK-SK,MINOS,ICARUS} (see Eq.(\ref{phase})).
Let us consider scheme A for definiteness. Then 
in vacuum the probabilities of
$ \nu_\alpha \to \nu_\beta $
transitions
in LBL experiments
are given by
\begin{equation}
P^{(\mathrm{LBL,A})}_{\nu_\alpha\to\nu_\beta}
=
\left|
U_{\beta1}
\,
U_{\alpha1}^{*}
+
U_{\beta2}
\,
U_{\alpha2}^{*}
\,
\exp\!\left(
- i
\frac{ \Delta{m}^{2}_{21} \, L }{ 2 \, p }
\right)
\right|^2
+
\left|
\sum_{k=3,4}
U_{{\beta}k}
\,
U_{{\alpha}k}^{*}
\right|^2
\,.
\label{plba}
\end{equation}
This formula has been obtained from Eq.(\ref{prob})
taking into account the fact that in LBL experiments
$ \Delta{m}^{2}_{43} L / 2 p \ll 1 $
and dropping the terms proportional to 
the cosines of phases much larger
than $2\pi$
($ \Delta{m}^{2}_{kj} L / 2 p \gg 2\pi $
for $k=3,4$ and $j=1,2$).
Such terms do not contribute to the oscillation
probabilities averaged over the
neutrino energy spectrum.

To obtain limits on the
LBL oscillation probability (\ref{plba})
from the results of the
SBL oscillation experiments, we employ
the Cauchy--Schwarz inequality
on the term with the summation over $k=1,2$ and use $c_\alpha$ 
(\ref{ca}) with $r=2$
to find the inequalites
\begin{equation}
\left( 1 - c_{\alpha} \right)^2 \leq
P^{(\mathrm{LBL,A})}_{\stackrel{\makebox[0pt][l]
{$\hskip-3pt\scriptscriptstyle(-)$}}{\nu_{\alpha}}
\to\stackrel{\makebox[0pt][l]
{$\hskip-3pt\scriptscriptstyle(-)$}}{\nu_{\alpha}}}
\quad \mbox{and} \quad
c^2_{\alpha} \leq
P^{(\mathrm{LBL,B})}_{\stackrel{\makebox[0pt][l]
{$\hskip-3pt\scriptscriptstyle(-)$}}{\nu_{\alpha}}
\to\stackrel{\makebox[0pt][l]
{$\hskip-3pt\scriptscriptstyle(-)$}}{\nu_{\alpha}}}
\label{paa}
\end{equation}
and
\begin{equation}
P^{(\mathrm{LBL})}_{\stackrel{\makebox[0pt][l]
{$\hskip-3pt\scriptscriptstyle(-)$}}{\nu_{\alpha}}
\to\stackrel{\makebox[0pt][l]
{$\hskip-3pt\scriptscriptstyle(-)$}}{\nu_{\beta}}}
\leq
c_{\alpha}
\,
c_{\beta}
+
\frac{1}{4}
\,
A_{\alpha;\beta} \quad (\alpha \neq \beta)
\;.
\label{pab1}
\end{equation}
It can easily be shown \cite{BGG97a} that 
Eq.(\ref{pab1}) is scheme-independent and that both equations also hold for
antineutrinos. Considering reactor experiments and 
taking into account Eq.(\ref{schemes})
we obtain the bound
\begin{equation}\label{ee}
1-P^{(\mathrm{LBL})}_{\bar\nu_e\to\bar\nu_e}
\leq a^0_e\, (2-a^0_e) 
\end{equation}
which holds for both schemes. Inserting the numerical values of the
function $a^0_e$ (see Fig.\ref{a0plot}) it turns out that the upper
bound (\ref{ee}) is below the sensitivity of the CHOOZ experiment in
the preferred range (\ref{LSNDrange}) of 
$\Delta m^2_{\mbox{\scriptsize SBL}}$. For the accelerator experiments
matter effects have to be taken into account. We have shown
\cite{BGG97a} that the
matter-corrected version of Eq.(\ref{pab1}) leads to stringent bounds
on 
$\stackrel{\scriptscriptstyle (-)}{\nu}_{\hskip-3pt \mu} 
\to
\stackrel{\scriptscriptstyle (-)}{\nu}_{\hskip-3pt e}$
and
$\stackrel{\scriptscriptstyle (-)}{\nu}_{\hskip-3pt e}
\to
\stackrel{\scriptscriptstyle (-)}{\nu}_{\hskip-3pt \tau}$
LBL transition probabilities of the order of $10^{-2}$ to $10^{-1}$ depending
on the value of $\Delta m^2_{\mbox{\scriptsize SBL}}$ and on the energy
of the neutrino beam (for a study of LBL CP violation in schemes A and
B see Ref.\cite{BGG97b}).

\section{Conclusions}

In this report we have discussed the possible form of the
neutrino mass spectrum
that can be inferred from
the results of all
neutrino oscillation experiments,
including the solar and atmospheric neutrino
experiments. The crucial input are the three indications in favour of
neutrino oscillations given by the solar neutrino data, the
atmospheric neutrino anomaly and the result of the LSND
experiment. These indications, which all pertain to different scales
of neutrino mass-squared differences, require that apart from the 
three well-know neutrino flavours at least one additional sterile
neutrino (without couplings to the $W$ and $Z$ bosons) must exist.
In our investigation we have
assumed that there is one sterile neutrino and that the
4-neutrino mixing matrix (\ref{mixing}) is unitary.
We have considered all possible
schemes with four massive neutrinos
which provide 
three scales of $\delta{m}^2$.
We have argued that a neutrino mass hierarchy is not compatible with
the above-mentioned indications in favour of
neutrino oscillations together with the negative results of all other
SBL neutrino oscillation experiments other than LSND.
The same holds for all mass spectra with three squares 
of neutrino masses
clustered together, such that the gap between the cluster and the
remaining mass-squared determines $\Delta m^2_{\mbox{\scriptsize SBL}}$ 
relevant in SBL experiments.

Thus only two possible spectra
of neutrino masses,
denoted by A and B
(see Eq.(\ref{AB})),
with two pairs of close
masses separated by a mass difference
of the order of 1 eV
are compatible with the results of
all neutrino oscillation experiments. The positive result of the LSND
experiment confines the SBL mass-squared to the interval
$0.27 \, \mathrm{eV}^2
\lesssim
\Delta{m}^2_{\mathrm{SBL}}
\lesssim
2.2 \, \mathrm{eV}^2$ (see Fig.\ref{emu}).
If, of the two neutrino schemes defined by Eqs.(\ref{AB}) and
(\ref{schemes}), scheme A
is realized in nature,
the neutrino mass that is measured
in $^3$H $\beta$-decay experiments
coincides with the
``LSND mass''.
If the massive neutrinos are
Majorana particles,
in the case of scheme A,
the experiments on the search for
$(\beta\beta)_{0\nu}$ decay
have good chances to
obtain a positive result. 
Furthermore, schemes A and B have severe consequences for
long-baseline neutrino oscillations: the
$\stackrel{\scriptscriptstyle (-)}{\nu}_{\hskip-3pt e}$ 
survivial probability is close to one and the
$\stackrel{\scriptscriptstyle (-)}{\nu}_{\hskip-3pt \mu} 
\to
\stackrel{\scriptscriptstyle (-)}{\nu}_{\hskip-3pt e}
$
and 
$\stackrel{\scriptscriptstyle (-)}{\nu}_{\hskip-3pt e} 
\to
\stackrel{\scriptscriptstyle (-)}{\nu}_{\hskip-3pt \tau}$ 
transitions are strongly constrained.
 
Finally, we can
ask ourselves what happens if not all experimental input data
leading to schemes A and B are confirmed in future
experiments. Among the many questions in this context, the two most
burning ones concern LSND and the zenith angle variation in the
atmospheric neutrino flux. 
Clearly, if LSND is not confirmed, three neutrinos are
sufficient. If one nevertheless requires a 4th neutrino with a mass in the
eV range for cosmological reasons then the neutrino spectrum is
likely to be hierarchial because region III (see Fig.\ref{square})
cannot be excluded in this case. If, on the other hand, the zenith
angle variation in the 
atmospheric neutrino flux is not confirmed, a 
3-neutrino mixing scheme with
$\Delta m^2_{\mbox{\scriptsize SBL}} \equiv
\Delta m^2_{\mbox{\scriptsize atm}} \sim 0.3$ eV$^2$ and other
definite predictions is possible \cite{CF}.
We have to wait for future experimental results to see if the present
interesting and puzzling situation concerning the neutrino mass and
mixing pattern persists.

\begin{flushleft}
\Large \textbf{Acknowledgement}
\end{flushleft}
W.G. would like to thank the organizers of the workshop for their
great hospitality and the stimulating and pleasant atmosphere.


\begin{thebibliography}{99}

\bibitem{gauge}
R.N. Mohapatra and P.B. Pal,
\textit{``Massive Neutrinos in Physics and
Astrophysics''},
World Scientific Lecture Notes in Physics, Vol. 41, World
Scientific, Singapore, 1991.

\bibitem{BP78-87}
B. Pontecorvo, Sov. Phys. JETP \textbf{26}, 984 (1968);
S.M. Bilenky and B. Pontecorvo,
Phys. Rep. \textbf{41}, 225 (1978);
S.M. Bilenky and S.T. Petcov,
Rev. Mod. Phys. \textbf{59}, 671 (1987).

\bibitem{osc}
C. Giunti at al., Phys. Rev. D \textbf{48}, 4310 (1993);
K. Kiers, S. Nussinov and N. Weiss, Phys. Rev. D \textbf{53},
537 (1996);
W. Grimus and P. Stockinger, Phys. Rev. D \textbf{54}, 3414 (1996);
C. Giunti and C.W. Kim, hep-ph/9711363;
H. Burkhardt \textit{et al.}, hep-ph/9803365;
C.W. Kim and A. Pevsner, \textit{``Neutrinos in Physics and
Astrophysics''}, Contemporary Concepts in Physics, Vol. 8,
Harwood Academic Press, Chur, Switzerland, 1993.

\bibitem{Homestake}
B.T. Cleveland \textit{et al.},
Nucl. Phys. B (Proc. Suppl.) \textbf{38}, 47 (1995).

\bibitem{Kam-sun}
K.S. Hirata \textit{et al.},
Phys. Rev. D \textbf{44}, 2241 (1991).

\bibitem{GALLEX}
GALLEX Coll.,
W. Hampel \textit{et al.},
Phys. Lett. B \textbf{388}, 384 (1996).

\bibitem{SAGE}
SAGE Coll., J.N. Abdurashitov \textit{et al.},
Phys. Rev. Lett. \textbf{77}, 4708 (1996).

\bibitem{SK-sun}
K. Inoue,
Talk presented at the
$5^{\mathrm{th}}$
International Workshop on
\textit{Topics in Astroparticle and Underground Physics},
Gran Sasso, Italy, September 1997 \\
(http://\-www-sk.\-icrr.\-u-tokyo.\-ac.\-jp/\-doc/\-sk/\-pub/\-pub\_sk.\-html);
R. Svoboda,
Talk presented at the Conference on
\textit{Solar Neutrinos: News About SNUs},
2--6 December 1997, Santa Barbara, California 
(http://\-www.\-itp.\-ucsb.\-edu/\-online/\-snu/).

\bibitem{SK}
M. Nakahata,
Talk presented at the
APCTP Workshop:
\textit{Pacific Particle Physics Phenomenology},
Seoul, Korea, October 31 -- November 2, 1997\\
(http://\-www-sk.\-icrr.\-u-tokyo.\-ac.\-jp/\-doc/\-sk/\-pub/\-pub\_sk.\-html).

\bibitem{Kam-atm}
Y. Fukuda \textit{et al.},
Phys. Lett. B \textbf{335}, 237 (1994).

\bibitem{IMB}
R. Becker-Szendy \textit{et al.},
Nucl. Phys. B (Proc. Suppl.) \textbf{38}, 331 (1995).

\bibitem{Soudan}
W.W.M. Allison \textit{et al.},
Phys. Lett. B \textbf{391}, 491 (1997).

\bibitem{SK-atm}
Super-Kamiokande Coll.,
Y. Fukuda \textit{et al.},
ICRR-Report-411-98-7 (hep-ex/9803006).

\bibitem{LSND}
C. Athanassopoulos \textit{et al.},
Phys. Rev. Lett. \textbf{77}, 3082 (1996).

\bibitem{HL97}
N. Hata and P. Langacker,
Phys. Rev. D \textbf{56}, 6107 (1997).

\bibitem{FLM97}
G.L. Fogli, E. Lisi and D. Montanino, hep-ph/9709473.

\bibitem{Valencia}
M.C. Gonzalez-Garcia \textit{et al.}, hep-ph/9801368.

\bibitem{MSW} S.P. Mikheyev and A.Yu. Smirnov,
Yad. Fiz. \textbf{42}, 1441 (1985)
[Sov. J. Nucl. Phys. \textbf{42}, 913 (1985)];
Il Nuovo Cimento C \textbf{9}, 17 (1986);
L. Wolfenstein,
Phys. Rev. D \textbf{17}, 2369 (1978);
\textit{ibid.} \textbf{20}, 2634 (1979).

\bibitem{four}
J.T. Peltoniemi and J.W.F. Valle,
Nucl. Phys. B \textbf{406}, 409 (1993);
D.O. Caldwell and R.N. Mohapatra,
Phys. Rev. D \textbf{48}, 3259 (1993);
Z. Berezhiani and R.N. Mohapatra,
Phys. Rev. D \textbf{52}, 6607 (1995);
J.R. Primack \textit{et al.},
Phys. Rev. Lett. \textbf{74}, 2160 (1995);
E. Ma and P. Roy,
Phys. Rev. D \textbf{52}, R4780 (1995);
R. Foot and R.R. Volkas,
Phys. Rev. D \textbf{52}, 6595 (1995);
E.J. Chun \textit{et al.},
Phys. Lett. B \textbf{357}, 608 (1995);
J.J. Gomez-Cadenas and M.C. Gonzalez-Garcia,
Z. Phys. C \textbf{71}, 443 (1996);
S. Goswami,
Phys. Rev. D \textbf{55}, 2931 (1997);
A.Yu. Smirnov and M. Tanimoto,
Phys. Rev. D \textbf{55}, 1665 (1997);
E. Ma,
Mod. Phys. Lett. A \textbf{11}, 1893 (1996);
E.J. Chun, C.W. Kim and U.W. Lee, hep-ph/9802209.

\bibitem{BGKP}
S.M. Bilenky, C. Giunti, C.W. Kim and S.T. Petcov,
Phys. Rev. D \textbf{54}, 4432 (1996).

\bibitem{BGG96}
S.M. Bilenky, C. Giunti and W. Grimus,
Proc. of
\textit{Neutrino '96}, Helsinki, June 1996, edited by K. Enqvist
\textit{et al.}, p.174 (World Scientific, Singapore, 1997);
Eur. Phys. J. C \textbf{1}, 247 (1998).

\bibitem{OY96} N. Okada and O. Yasuda, 
Int. J. Mod. Phys. A \textbf{12}, 3669 (1997).

\bibitem{BGG97a}
S.M. Bilenky, C. Giunti and W. Grimus,
Phys. Rev. D \textbf{57} (1998) 1920.

\bibitem{BGG97b}
S.M. Bilenky, C. Giunti and W. Grimus, hep-ph/9712537, 
to be published in Phys. Rev. D.

\bibitem{Barger-Gibbons}
V. Barger, T. Weiler and K. Whisnant, hep-ph/9712495;
S.C. Gibbons, R.N. Mohapatra, S. Nandi and A. Raychaudhuri,
hep-ph/9803299.

\bibitem{Bugey95}
B. Achkar \textit{et al.},
Nucl. Phys. B \textbf{434}, 503 (1995).

\bibitem{CDHS84}
F. Dydak \textit{et al.},
Phys. Lett. B \textbf{134}, 281 (1984).

\bibitem{CCFR84}
I.E. Stockdale \textit{et al.},
Phys. Rev. Lett. \textbf{52}, 1384 (1984).

\bibitem{BBGK}
S.M. Bilenky, A. Bottino, C. Giunti and C.W. Kim,
Phys. Lett. B \textbf{356}, 273 (1995);
Phys. Rev. D \textbf{54}, 1881 (1996).

\bibitem{BNLE734}
L.A. Ahrens \text{et al.}, Phys. Rev. D \textbf{36}, 702 (1987).

\bibitem{BNLE776}
L. Borodovsky \textit{et al.},
Phys. Rev. Lett. \textbf{68}, 274 (1992).

\bibitem{KARMEN}
J. Kleinfeller,
Nucl. Phys. B (Proc. Suppl.) \textbf{48}, 207 (1996).

\bibitem{CCFR97}
A. Romosan \textit{et al.}, Phys. Rev. Lett. \textbf{78}, 2912 (1997).

\bibitem{SS92}
X. Shi and D.N. Schramm,
Phys. Lett. B \textbf{283}, 305 (1992).

\bibitem{kearns}
E. Kearns, Talk presented at the Conference on
\textit{Solar Neutrinos: News About SNUs},
2--6 December 1997, Santa Barbara, California 
(http://\-www.\-itp.\-ucsb.\-edu/\-online/\-snu/).

\bibitem{KP96}
P.I. Krastev and S.T. Petcov,
Phys. Rev. D \textbf{53}, 1665 (1996).

\bibitem{CHOOZ}
M. Appolonio \textit{et al.}, hep-ex/9711002.

\bibitem{PV}
F. Boehm \textit{et al.},
\textit{The Palo Verde experiment},
1996\\
(http://www.cco.caltech.edu/~songhoon/Palo-Verde/Palo-Verde.html).

\bibitem{KEK-SK}
Y. Suzuki,
Proc. of
\textit{Neutrino '96}, Helsinki, June 1996, edited by K. Enqvist
\textit{et al.}, p.237 (World Scientific, Singapore, 1997).

\bibitem{MINOS}
MINOS Coll.,
NUMI-L-63, February 1995.

\bibitem{ICARUS}
ICARUS Coll.,
LNGS-94/99-I,
May 1994.

\bibitem{CF}
C.Y. Cardall and G.M. Fuller, Phys. Rev. D \textbf{53}, 4421 (1996);
C.Y. Cardall, G.M. Fuller and D.B. Cline, 
Phys.Lett. B \textbf{413}, 246 (1997);
E. Ma and P. Roy, hep-ph/9706309.

\end{thebibliography}
\end{document}